\documentclass[12pt,a4paper]{article}

\pdfoutput=1

\usepackage[utf8x]{inputenc}
\usepackage[T1]{fontenc}
\usepackage{lmodern}
\usepackage[english]{babel}

\usepackage[rmargin=2.5cm,
            lmargin=2.5cm,
            top=2.5cm,
            bottom=2.5cm]{geometry}
            
\usepackage{enumerate}
\usepackage{paralist}
\renewcommand{\labelenumi}{(\theenumi)}

\usepackage{graphicx}
\usepackage{subfig}
\usepackage{placeins}

\usepackage{amsmath}
\usepackage{amsfonts}
\usepackage{amssymb}
\usepackage{dsfont}

\usepackage{longtable}
\usepackage{multirow}
\usepackage{rotating}

\usepackage[colorlinks=false, pdfborder={0 0 0}]{hyperref}

\usepackage{natbib}
\usepackage{url}

\usepackage{caption}
\usepackage{framed}

\usepackage{color}

\newcommand{\vectheta}{\boldsymbol{\theta}}

\newcommand{\veclambda}{\boldsymbol{\lambda}}
\newcommand{\vecbeta}{\boldsymbol{\beta}}
\newcommand{\vecu}{\boldsymbol{u}}
\newcommand{\vecB}{\boldsymbol{B}}
\newcommand{\matA}{\boldsymbol{A}}
\newcommand{\vecb}{\boldsymbol{b}}
\newcommand{\vecd}{\boldsymbol{d}}
\newcommand{\matD}{\boldsymbol{D}}
\newcommand{\matF}{\boldsymbol{F}}
\newcommand{\matH}{\boldsymbol{H}}
\newcommand{\vecs}{\boldsymbol{s}}

\newcommand{\by}{\boldsymbol{y}}
\newcommand{\bY}{\boldsymbol{Y}}

\newcommand{\Prob}{\mathbb{P}}
\newcommand{\E}{\mathbb{E}}
\newcommand{\Var}{\mathrm{Var}}

\newcommand\independent{\protect\mathpalette{\protect\independenT}{\perp}}
\def\independenT#1#2{\mathrel{\rlap{$#1#2$}\mkern2mu{#1#2}}}

\newcommand{\Lik}{\mathcal{L}}
\newcommand{\lik}{\ell}

\begin{document}

\begin{center}

{\Large \bfseries Stable Exponential Random Graph Models with Non-parametric Components for Large Dense Networks}
\vspace{5 mm}

{\large S. Thiemichen$^a$, G. Kauermann$^a$. } \\
{\textit{$^a$Institut f\"ur Statistik  Ludwigs-Maximilians-Universit\"at M\"unchen, Germany.}}

\vspace*{5mm}

\today 

\vspace{5mm}

\end{center}

\begin{abstract}
Exponential Random Graph Models (ERGM) behave peculiar in large networks with thousand(s) of actors (nodes). Standard models containing two-star or triangle counts as statistics are often unstable leading to completely full or empty networks. Moreover, numerical methods break down which makes it complicated to apply ERGMs to large networks. In this paper we propose two strategies to circumvent these obstacles. First, we fit a model to a subsampled network and secondly, we show how linear statistics (like two-stars etc.) can be replaced by smooth functional components. These two steps in combination allow to fit stable models to large network data, which is illustrated by a data example including a residual analysis.
\end{abstract}

\textbf{Keywords:} Exponential random graph models; Conditional independence; Subsampling; Smooth non-parametric components; Network analysis

\newpage

\section{Introduction}
\label{sec:Introduction}

The analysis of network data is an emerging field in statistics. It is challenging both model-wise and computationally. Recently, \cite{GoldenbergZheng:2010}, \cite{HunterKrivitskySchweinberger:2012}, and \cite{Fienberg:2012} published comprehensive survey articles discussing new statistical approaches and developments in network data analysis. We also refer to the monograph of \citet{Kolaczyk:2009} for a general introduction to the field, or the recent book of \cite{LusherKoskinenRobins:2013}, which focuses on a specific and widely used class of network models, so-called Exponential Random Graph Models (ERGM).\\

In its most simple form a network consists of a set of $n$ nodes (actors) which are potentially linked with each other through edges. These edges between the actors are thereby the focus of interest. Notationally a network can be expressed as a $n \times n$ (random) adjacency matrix $\bY$ with entries $Y_{ij} = 1$ if node $i$ and $j$ are connected, and $Y_{ij} = 0$ otherwise. In undirected networks one has $Y_{ij} = Y_{ji}$ while for directed links we have $Y_{ij} = 1$ if a directed edge goes from node $i$ to node $j$. For the sake of readability and notional simplicity we will concentrate here on undirected networks. The term $\by$ denotes a concrete realisation of $\bY$.\\

A common and powerful model for network data $\bY$ was proposed by \citet{FrankStrauss:1986} as Exponential Random Graph Model (ERGM) taking the form
\begin{align}\label{eq:ERGM}
\Prob\bigl( \bY = \by | \vectheta \bigr) = \frac{ \exp\left\{ \sum \limits_{l = 0}^p s_l(\by)\theta_l\right\}}{\kappa(\vectheta)},
\end{align}
with $\vectheta = (\theta_0, \ldots, \theta_p)^t$ as parameter vector and $s(\by) = (s_0(\by), \ldots, s_p(\by))^t$ as vector of statistics of the network. In equation (\ref{eq:ERGM}) the term $\kappa(\vectheta)$ denotes the normalizing constant, that is
\[
\kappa(\vectheta) = \sum\limits_{\by \in \mathcal{Y}} \exp\left\{\vectheta^t s(\by) \right\},
\]
where $\mathcal{Y}$ is the set of all networks and accordingly the sum is over $2^{\binom{n}{2}}$ terms. It is therefore numerically intractable, except for very small graphs. We denote with $s_0(\by)~=~\sum_{i = 1}^n~\sum_{j > i}^n~y_{ij}$ the baseline~statistic giving the number of edges in the (undirected) network, so that $\theta_0$ serves as intercept. The interpretation of the remaining parameters $\theta_l,\ l = 1, \ldots, p$, results through the corresponding conditional model for each single edge $Y_{ij}$ given the remaining network $\bY \backslash Y_{ij}$, since
\begin{align}\label{eq:condERGM}
\text{logit}\left[ \Prob\bigl( Y_{ij} = 1 | \bY \backslash Y_{ij} ; \vectheta\bigr) \right] = \theta_0 + \sum \limits_{l = 1}^{p} \Delta_{ij} s_l(\by) \theta_l,
\end{align}
where $\Delta_{ij} s_l(\by) = s_l(\by \backslash y_{ij}, y_{ij} = 1) - s_l(\by \backslash y_{ij}, y_{ij} = 0)$ is the so-called change statistics which is obtained by flipping the edge between nodes $i$ and $j$ from non-existent to existent.\\

Exponential Random Graph Models are numerically unstable, in particular if the number of actors $n$ gets large. Hence, for large networks one is faced with two relevant problems. First, the model itself is notoriously unstable leading to either full or empty networks. This issue is usually called degeneracy problem, see, for example, \citep{Schweinberger:2011}, and \cite{ChatterjeeDiaconis:2013}. Secondly, the estimation is per se numerically demanding or even unfeasible since numerical simulation routines are too time consuming. We aim to tackle both problems in this paper. First, we propose the use of stable statistics which are derived as smooth, non-parametric curves. Secondly, instead of fitting the model to the entire network we propose to draw samples from the network such that estimation in each sample is numerically (very) easy. These two proposals allow to easily analyse network data in large and sufficiently dense networks.\\

\cite{Schweinberger:2011} denotes network statistics (and the corresponding ERGM) as unstable if the statistics is not at least of order $O_p(n)$. In fact he shows that any $k$-star or triangle statistics is unstable leading to an odd behaviour of model (\ref{eq:ERGM}). Effectively, unstable networks are either complete (i.e. have all possible edges) or empty (i.e. all nodes are unconnected) unless for a diminishing subspace of the parameter space for $n$ increasing. If $n$ gets large it is therefore advisable to replace the statistics in model (\ref{eq:ERGM}) by stable statistics of order $O_p(n)$. A first proposal in this direction are alternating star and alternating triangle statistics as proposed in \cite{Snijders-etal:2006}, or geometrically weighted statistics as proposed in the context of Curved Exponential Random Graph Models, see \cite{HunterHandcock:2006}. \cite{Hunter:2007} shows that from a modelling point of view the alternating statistics are equivalent to geometrically weighted degree or geometrically weighted edgewise shared partners, respectively. Both approaches stabilize the models but for the price of less intuitive interpretations of the parameter estimates. We propose an alternative by making use of non-parametric models based and the technique of smoothing (see, e.g., \citealp{Ruppert-etal:2003}). The non-parametric model thereby maintains the interpretability of the ERGM based on the conditional model (\ref{eq:condERGM}). To motivate our idea we start with the conditional model (\ref{eq:condERGM}) and replace the linear terms through non-linear smooth components. This leads to the conditional non-parametric model
\begin{align}\label{eq:condsmooth}
\text{logit}\bigl[ \Prob\bigl( Y_{ij} = 1 | \bY \backslash Y_{ij}\bigr) \bigr] = \theta_0 + \sum \limits_{l = 1}^{p} m_l(\Delta_{ij} s_l(\by)),
\end{align}
where $m_l(\cdot)$ are smooth functions which need to be estimated from the data. Models of type (\ref{eq:condsmooth}) have been proposed in a simple regression framework as generalized additive models, see, e.g., \cite{HastieTibshirani:1990}, or \cite{Wood:2006}, but apparently the structure here is more complex as we are tackling network data. We additionally need to postulate that functions $m_l(\cdot)$ are monotone and bounded which in turn leads to stable network statistics in the definition of \cite{Schweinberger:2011}. We make use of penalized spline smoothing which also allows to accommodate constraints on the functional shape leading to stable network models. In fact, assuming $m_l(\cdot)$ to be monotone and bounded, we may derive a non-parametric Exponential Random Graph Model from (\ref{eq:condsmooth}) which takes the form
\begin{align}\label{eq:NPERGM}
\Prob \bigl( \bY = \by | \theta_0, m_l(\cdot), l = 1, \ldots, p \bigr) = \frac{ \exp\left\{s_0(\by)\theta_0 + \sum \limits_{l = 1}^p \sum \limits_i \sum \limits_{j>i} y_{ij} m_l(\Delta_{ij}s_l(\by)) \right\} }{ \kappa(\theta_0, m_l(\cdot), l = 1, \ldots, p) }
\end{align}
Apparently, model (\ref{eq:NPERGM}) appears rather complex due to its semi-parametric structure and estimation looks like a challenging task. We will argue, however, that smoothing techniques can easily be applied and estimation becomes feasible by making use of sampling strategies in networks leading to numerically simple likelihoods and in fact consistent (though not efficient) estimates.\\

\begin{figure}[htb!]\centering
\begin{tabular}{ccc}
\includegraphics[width=0.35\textwidth]{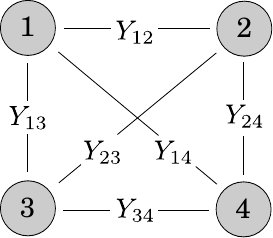} & \hspace*{1.5cm} &\includegraphics[width=0.3\textwidth]{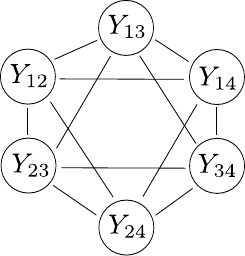} \\
\end{tabular}
\caption{Visualisation of the induced Markov independence graph (right) for a Exponential Random Graph Model for a simple 4-node network (left).}\label{fig:inducedIndependence}
\end{figure}

Estimation in Exponential Random Graph Models is cumbersome and numerically demanding as it requires simulation based routines. \cite{Snijders:2002} suggests the calculation of $\partial \kappa(\vectheta)/\partial \vectheta$ in the score equation resulting from (\ref{eq:ERGM}) using stochastic approximation. \cite{HunterHandcock:2006} propose to use MCMC methods in order to obtain the maximum likelihood estimate. The approach is extended and improved in \cite{HummelHunterHandcock:2012}. In a recent paper \cite{CaimoFriel:2011} develop a fully Bayesian estimation routine by incorporating the so-called exchange algorithm from \cite{Murray-etal:2006} which circumvents the calculation or approximation of the normalisation constant for the price of extended MCMC sampling. A general survey of available routines for fitting Exponential Random Graph Models is given in \citet{HunterKrivitskySchweinberger:2012}. In fact, if the network is large, MCMC based routines readily become numerically infeasible. As aforementioned, we will therefore make use of subsampling the network data and fit the model to subsamples that allow for simple likelihoods. We follow ideas of \cite{KoskinenDaraganova:2013}. In fact, for models with $k$-stars or triangles only, the edges follow a Markovian independence structure by conditioning on parts of the network (see \citealp{FrankStrauss:1986}, or \citealp{Whittaker:2009}). This is exemplified in a simple network with four nodes in Figure \ref{fig:inducedIndependence}. Conditioning on edges $Y_{12}, Y_{14},Y_{23}$, and $Y_{34}$ we find that $Y_{13}$ and $Y_{24}$ are conditionally independent, which can be denoted as $Y_{13} \independent Y_{24} | \bY \backslash \{Y_{13}, Y_{24}\}$. The idea is now to make use of this independence property to fit model (\ref{eq:NPERGM}) to a subsample of the network while conditioning on the rest of the network. Hence, exemplary we sample edges $Y_{13}$ and $Y_{24}$, and condition on $\bY \backslash \{Y_{13},Y_{24}\}$. Due to the (conditional) independence structure we can easily fit the conditional model (\ref{eq:condERGM}) with standard software for generalized linear and non-parametric additive models. This will be demonstrated below. Apparently such a strategy is not efficient if the network is small, but if the network is (very) large and (sufficiently) dense, sampling appears as a plausible approach which also maintains numerical feasibility.\\

The paper is organized as follows. In Section \ref{sec:estimation} we suggest to estimate large Exponential Random Graph Models through subsampling of the network. In Section \ref{sec:nonparametric} we extend the idea towards non-parametric models. Section \ref{sec:data} gives a data example demonstrating the usability of the approach. Finally, a discussion completes the paper in Section \ref{sec:discussion}. \\
All routines for fitting and analysing the models are written in \texttt{R} \citep{RCore:2016} and will be made available as \texttt{R} package.

\section{Estimation through Subsampling}
\label{sec:estimation}


The general idea proposed in this section is that instead of fitting an ERGM to the entire data, we fit a conditional model to appropriate subsamples of the data. Due to conditional independence this allows for fast and easy computing. We start the presentation with the classical (unstable) ERGM and assume model (\ref{eq:ERGM}) has statistics like $k$-star and triangle effects only. That is statistics $s_l(\cdot)$ in (\ref{eq:ERGM}) for instance has \underline{no} ``4-cycles''$~$of the form $ \sum\nolimits_{i<j<k<l} Y_{ij} Y_{jk} Y_{kl} Y_{li} $ (or higher order cycles). Let us get more specific. For simplicity of presentation let $n$, the number of nodes in the network, be even. With $\mathcal{D}(n|2)$ we denote a decomposition of the set $\{1, \dots, n\}$ into subsets of size 2, e.g.,\ $\mathcal{D}(n|2) = \{(1,2), (3,4), \dots, (n-1, n)\}$. For $A = (i,j) \in \mathcal{D}(n|2)$ we denote $Y_A = Y_{ij}$ and $\bY \backslash Y_{\mathcal{D}(n|2)} = \{Y_{ij}, (i,j) \notin \mathcal{D}(n|2)\}$. Apparently $\mathcal{D}(n|2)$ has $n/2$ elements. We assume now that the statistics $s_l(\by)$ in \eqref{eq:ERGM} can be decomposed to
\begin{align}
s_l(\by) = \sum_{A \in \mathcal{D}(n|2)} s_{lA} (y_A, \by \backslash y_{\mathcal{D}(n|2)}).
\label{eq:condstat}
\end{align} 
This holds for all $k$-stars and triangle statistics. It is not difficult to show that with condition (\ref{eq:condstat}) density (\ref{eq:ERGM}) can then be factorized to
\begin{align}\label{eq:fac}
\Prob\bigl( \bY = \by \bigr) = \prod_{A \in \mathcal{D}(n|2)} h_A( y_A, \by \backslash y_{\mathcal{D}(n|2)}),
\end{align} 
where $h_A(\cdot)$ is some function depending on $\{ y_A, \by \backslash y_{\mathcal{D}(n|2)} \}$. The factorization \eqref{eq:fac} implies that the edges with indices in $\mathcal{D}(n|2)$ are mutually independent conditional on the rest of the network (see \citealp{Whittaker:2009}).\\
The conditional independence will be used to fit model \eqref{eq:ERGM} not for the entire network but for an appropriately chosen subnetwork. We therefore draw a sample of the network $\bY$ by taking $\by_A$ with $A \in \mathcal{D}(n|2)$ as sampled binary observations accompanied by $\Delta_A s(\by) = (\Delta_A s_1(\by), ..., \Delta_A s_p(\by))^t$ as corresponding change statistics with obvious definition of $\Delta_A s(\cdot)$. The term $\Delta_A s(\by)$ plays the role of covariates and the conditional model (\ref{eq:condERGM}) takes the form
\begin{align*}
\text{logit} \left[\Prob\bigl( Y_A = 1 | \Delta s(\by),\vectheta \bigr) \right] =\theta_0 + \sum \limits_{l=1}^{p} \Delta_A s_l (\by) \theta_l = \theta_0 + \sum \limits_{l=1}^{p} x_l \theta_l,
\end{align*}
where $x_l$ denotes the change statistics $\Delta_A s_l (\by)$ which is considered as covariate in the logit model. Due to the induced conditional independence the likelihood for the sample results to
\begin{align}\label{eq:lik}
\Lik_{\mathcal{D}(n|2)}( \vectheta ) = \prod \nolimits_{A \in \mathcal{D}(n|2)} \Prob\bigl( Y_A = 1 | \Delta_A s(\by), \vectheta\bigr),
\end{align}
which is easily fitted using standard software for generalized linear models. Note that (\ref{eq:lik}) is the true likelihood for the conditional subsample so that consistent estimates and their variance estimates are easily available. This means, by taking the subsample of edge variables $Y_A$ with $A \in \mathcal{D}(n|2)$ and conditioning in the remaining graph we circumvent numerical estimation problems and remain in the classical generalized linear model framework. It also implies that we can estimate $\vectheta$ consistently (for $n$ increasing) by maximizing $\Lik_{\mathcal{D}(n|2)}(\vectheta)$.\\

\begin{figure}[htb!]
\begin{center}
\includegraphics[width=0.35\textwidth]{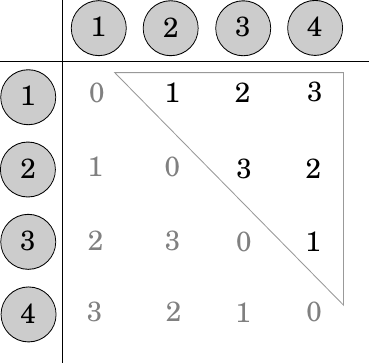}
\caption{\label{fig:ls}Symmetric Latin Square with unique diagonal.}
\end{center}
\end{figure}

Apparently, we may draw different samples of edges leading to different estimates. This means using different decomposition sets $\mathcal{D}(n|2)$ leads to different estimates. This leaves us with the question how to combine the different estimates. We may either draw $\mathcal{D}(n|2)$ randomly or make use of a combinatorial approach to cover the entire network $Y_{ij}$. Let therefore 
\[
\mathcal{P} = \{ \mathcal{D}_k(n|2), k = 1, \ldots, n-1 \}
\] 
be a sequence of sets $\mathcal{D}_k(n|2)$ such that each index pair is exactly in one single set $\mathcal{D}_k(n|2)$. That is for $Y_{ij}$ there exists exactly one set $\mathcal{D}_k(n|2) \in \mathcal{P}$ with $(i, j) \in \mathcal{D}_k(n|2)$. The $n-1$ sets $\mathcal{D}_k(n|2)$ in $\mathcal{P}$ can be constructed using a symmetric Latin Square with a unique diagonal (see, e.g.,\  \citealp{AndersenHilton:1980}).\footnote{A description of a possible algorithm for the construction of such a symmetric Latin square with a unique diagonal is available, e.g., from \cite{LS}.} For instance for $n = 4$ nodes Figure \ref{fig:ls} shows a symmetric Latin Square. As we are focusing on undirected networks, where the corresponding adjacency matrix is symmetric, we use only the upper diagonal of the Latin Square. We may take the entries in the Latin Square as the sample number. For instance, $\mathcal{D}_1(n|2) $ results by taking the pairs with entries 1 in the upper triangle from the corresponding network adjacency matrix, i.e.\ $(1,2),(3,4)$ and condition on the remaining variables. Accordingly we proceed for entries $2$ and $3$ in the Latin Square. We denote with $\widehat{\vectheta}_{<k>}$ the resulting estimate from sequence set $\mathcal{D}_k(n|2)$. Note that each estimate $\widehat{\vectheta}_{<k>}$ is consistent but they are not mutually independent. 
With $Y_{<k>} = \{Y_{ij}:(i,j) \in \mathcal{D}_k(n|2)\}$ we easily get with the asymptotic properties of Maximum Likelihood estimates as $n \rightarrow \infty$ that
\begin{align*}
\E\bigl( \widehat{\vectheta}_{<k>} \bigr)= \E_{\bY \backslash Y_{<k>}} \left( \E_{Y_{<k>}} \bigl( \widehat{\vectheta}_{<k>} | \bY \backslash Y_{<k>} \bigr) \right) \rightarrow \vectheta.
\end{align*}
Moreover
\begin{align*}
\mathrm{Var}\bigl( \widehat{\vectheta}_{<k>} \bigr) 
&= \E_{\bY \backslash Y_{<k>}} \left( \Var_{Y_{<k>}} \bigl( \widehat{\vectheta}_{<k>} | \bY \backslash  Y_{<k>} \bigr)\right) \\
                  & \quad +\ \Var_{\bY \backslash Y_{<k>}} \left( \E_{Y_{<k>}} \bigl( \widehat{\vectheta}_{<k>} | \bY \backslash Y_{<k>} \bigr) \right) \nonumber\\
                  &\rightarrow \E_{\bY \backslash Y_{<k>}} \left( F^{-1}_{<k>} \bigl( \vectheta_{<k>} \bigr) \right), \nonumber
\end{align*}
where $F_{<k>} \bigl( \vectheta \bigr)$ denotes the (conditional) Fisher matrix corresponding to the likelihood function (\ref{eq:lik}). Apparently $F^{-1}_{<k>}\bigl(  \vectheta_{<k>} \bigr)$ is an unbiased estimate for $\E_{\bY \backslash Y_{<k>}} \left( F^{-1}_{<k>} \bigl( \vectheta_{<k>} \bigr) \right)$. Note that $F^{-1}_{<k>} \bigl( \widehat{\vectheta}_{<k>} \bigr)$ can be obtained with any software package for fitting logistic regression models. Hence an estimate for the variance is readily available.

\section{Non-parametric Exponential Random Graph Models}
\label{sec:nonparametric}

\subsection{Spline-Based Model}
\label{subsec:nonparametric}

We have shown how an appropriate sample of the network allows for simple estimation of the parameters. Apparently this is a recommendable approach only if $n$, the number of nodes, is large. In this case, however, ERGMs become unstable if the change statistics increase linearly in $n$. As shown in \cite{Schweinberger:2011} this holds for almost all basic models with ${\theta}_l \neq 0$ for $l > 0$. In other words, even though we are able to estimate the parameters as described before, the resulting network will be either full or empty as $n$ is becoming large. Stability is achieved if the network statistics are of order $O_p(n)$. One intention is therefore to modify the statistics in the model such that they become stable. This is done with non-parametric components so that the change statistics have a bounded influence. To do so we make use of the non-parametric model \eqref{eq:condsmooth} where we additionally postulate that the smooth functions $m_l(\cdot)$ are monotone and bounded.\\

\begin{figure}[htb!]\centering
\includegraphics[width=0.9\textwidth]{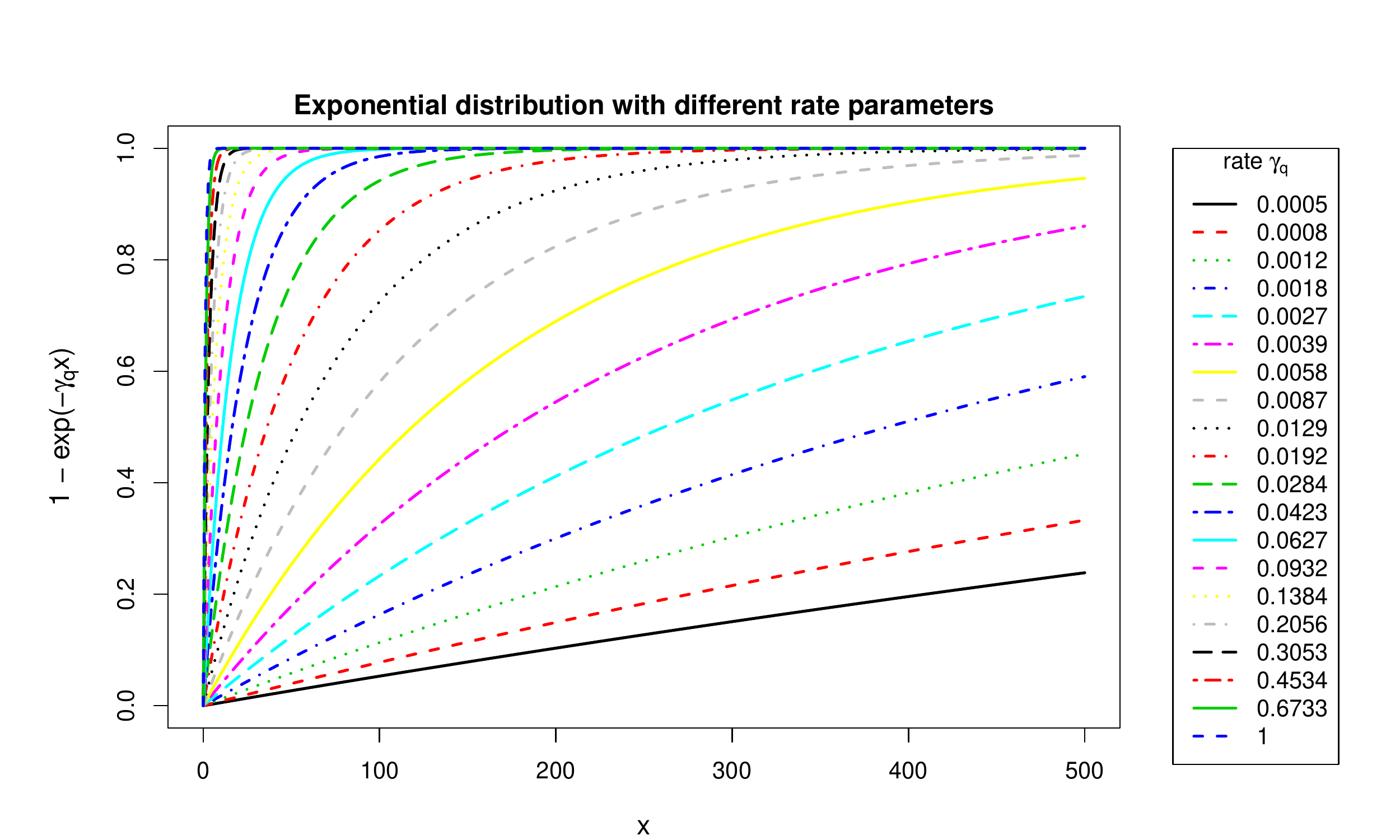}
\caption{Visualisation of the cumulative distribution function of the exponential distribution with different rate parameters $\gamma_q$ as example of possible basis functions $B_q(x)$.}\label{fig:basis}
\end{figure}

To estimate functions $m_l(\cdot)$ we make use of penalized spline smoothing as discussed in detail in \cite{Ruppert-etal:2003}, and \cite{Ruppert-etal:2009}, see also \cite{Kauermann-etal:2009}. The general idea is as follows. First, one replaces the unknown smooth function $m_l(\cdot)$ by a spline basis which is flexible (i.e. high dimensional) enough to capture the underlying true functional relation. As a second step a penalty or regularization is imposed on the unknown spline coefficients leading to a smooth and numerically stable fit. The third step is to calibrate/estimate the amount of penalization, which is controlled by a smoothing parameter. The original idea goes back to \cite{OSullivan:1986} and was made popular by the seminal paper of \cite{EilersMarx:1996}. We make use of the idea here, but amend it towards the specific problem of non-stability occurring in large networks. As first step we choose a basis $\vecB(x) = \left( B_1(x), \ldots, B_K(x) \right)^t$ where $x \in \mathbb{R}^+$ and the basis components $B_q(x)$, for $q = 1, \ldots, K$, fulfill the following three properties:
\begin{itemize}
\item[1)] $B_q(0) \equiv 0$,
\item[2)] $B_q(x)$ is monotone, and
\item[3)] $B_q(x)$ is bounded for $x \rightarrow \infty$.
\end{itemize}
A convenient choice are distribution functions on $\mathbb{R}^+$. Here we employ the exponential distribution and set
\begin{align}
B_q(x) = 1 - \exp(-\gamma_q x),
\end{align}
where $\gamma_q$ are fixed scaling parameters. The set $\{ \gamma_1, \ldots, \gamma_K\}$ covers a wide range of possible shapes as visualized in Figure \ref{fig:basis}. We now replace the unknown function $m_l(\cdot)$ in model \eqref{eq:condsmooth} by the spline representation 
\begin{align}
m_l(x_l) = \vecB(x_l)^t \vecu_l,
\label{eq:spline1}
\end{align}
with $\vecB(x) = \bigl( B_1(x), \ldots, B_K(x) \bigr)^t$ and $\vecu_l = (u_{l1}, \ldots, u_{lK} )^t$ as the coefficient vector. Note that as long as the coefficients of $\vecu_l$ are finite we have constructed a bounded and hence stable network statistics. Apparently we need additional constraints on $\vecu_l$ in order to guarantee monotonicity. This implies for monotonically increasing functions  that
\begin{equation}
\vecB'(x)^t \vecu_l \geq 0,
\label{eq:monoton}
\end{equation}
where $\boldsymbol{B}'(x)=\bigl( \gamma_1 \exp(- \gamma_1 x),..., \gamma_K \exp(-\gamma_K x)\bigr)^t$. This is a linear constraint on the parameters, which for estimation is easily accommodated by quadratic programming. For monotonically decreasing functions we use almost the same constraint. For practical purposes we select the cutpoints of neighbouring basis functions $\xi_r$ with $\gamma_{r+1} \exp (-\gamma_{r+1} \xi_r) = \gamma_r \exp(- \gamma_r \xi_r)$ and set the constraints to $\vecB'(\xi_r)^t \vecu_l \geq 0$ for monotonically increasing functions (or to $\vecB'(\xi_r)^t \vecu_l \leq 0$ for monotonically decreasing functions), for $r = 1, \ldots, K-1$. Our experiences show a stable behaviour with this setting.

\subsection{Penalized Estimation}
\label{subsec:pen}
The second step is now to impose a penalty on the spline coefficients in order to achieve smoothness and numerical stability. For a sample of the network as proposed in the previous section, let $\lik( \theta_0, \vecu )$ be the log-likelihood resulting from model \eqref{eq:condsmooth} in combination with \eqref{eq:spline1}, where $\vecu = \left( \vecu^t_1, \ldots, \vecu^t_p \right)^t$. For notational simplicity we omit the sampling index in this subsection. We emphasize however that the likelihood and hence its estimate do depend on the particular sample of the network. Bear in mind that $\vecu$ is high dimensional, so that (ML) estimates are unstable and the resulting fits $\vecB(x)\widehat{\vecu}_l$ would be wiggled. We therefore apply a ridge penalty leading to the penalized log-likelihood

\begin{align}
\lik_p\bigl( \theta_0, \vecu, \veclambda \bigr) = \lik\bigl( \theta_0, \vecu \bigr) - \frac{1}{2} \sum_{l=1}^p \lambda_l \vecu^t_l \vecu_l,
\label{eq:penlik}
\end{align}

where $\veclambda = \left( \lambda_1, \ldots, \lambda_p \right)^t$ are the penalty parameters. Apparently setting $\lambda_l \rightarrow \infty$ leads to $m_l(\cdot) \equiv 0$ while $\lambda_l \rightarrow 0$ gives an unpenalized fit. It remains therefore to choose $\veclambda$ data driven balancing goodness of fit ($\veclambda \rightarrow 0$) and parsimony of the model (minimal for $\veclambda \rightarrow \infty$). These steps can be carried out with classical cross-validation (see, e.g., \citealp{EilersMarx:1996}) or in a more sophisticated way by comprehending the penalty as normal prior. In this case we follow a Bayesian view and assume $\vecu_l \sim N\bigl( 0, \lambda_l^{-1} I_K \bigr)$ with $I_K$ as $K$ dimensional unit matrix. Then $\lambda_l$ is the reciprocal of the a priori variance of $\vecu_l$. The connection between penalized estimation and its Bayesian view by imposing normal priors is extensively motivated and discussed in \cite{Ruppert-etal:2003}. In fact the approach led to a real breakthrough in smooth functional estimation as mirrored in the survey article by \cite{Ruppert-etal:2009}. Note that the Bayesian approach in our setting here leads to a generalized linear mixed model which is extensively discussed, e.g., in \cite{BreslowClayton:1993}, see also \cite{McCulloch-etal:2008}. In particular, assuming a normal prior for coefficient vector $\vecu_l$ we may consider the penalty $\lambda_l$ as parameter which needs to be estimated. To do so we make use of the procedure of \cite{Schall:1991} leading to the following formulae. With $F(\theta_0, \vecu)$ we denote the Fisher matrix of the conditional model \eqref{eq:condERGM}. We define the Fisher matrix in the penalized likelihood \eqref{eq:penlik} as 
\begin{align}
\matF\bigl( \theta_0, \vecu, \veclambda \bigr)= \matF\bigl( \theta_0, \vecu \bigr) + \mbox{diag}\bigl( 0, \lambda_1 I_K, \ldots, \lambda_p I_K \bigr),
\end{align}
where $\mbox{diag}(\cdot)$ denotes a block diagonal matrix with the arguments as blocks. The part of the Fisher matrix belonging to $\vecu$ is then
\begin{align}
\widetilde{\matF} \bigl( \vecu, \veclambda \bigr) = \widetilde{\matF} \bigl( \vecu \bigr) + \mbox{diag}\bigl( \lambda_1 I_K, \ldots, \lambda_p I_K \bigr),
\end{align}
with $\widetilde{\matF}(\vecu)$ denoting the part of the Fisher matrix from the conditional model \eqref{eq:condERGM} belonging to $\vecu$. Following \cite{Schall:1991} we can now estimate $\lambda_l^{-1}$ (iteratively) through
\begin{align}\label{eq:schall_main}
\widehat{\lambda}_l^{-1} = \frac{\vecu_l^t \vecu_l}{\mbox{df} \bigl( \lambda_l \bigr)},
\end{align}
where 
\begin{align*}
\mbox{df}\bigl( \lambda_l \bigr) = \mbox{tr} \left\lbrace \bigl[ \widetilde{\matF}^{-1} \bigl( \vecu, \veclambda \bigr) \widetilde{\matF}\bigl( \vecu, 0 \bigr) \bigr]_l \right\rbrace,
\end{align*}
and subscript $l$ means that we take only the submatrix matching to component $\vecu_l$. See \cite{Kauermann:2005}, or \cite{KrivobokovaKauermann:2007} for a derivation of the estimate. Finally, the monotonicity constraint \eqref{eq:monoton} is taken into account by quadratic programming which is available in \texttt{R} using the package \texttt{quadprog} \citep{quadprog}. The following algorithm describes the iterative procedure.

\begin{framed}
\textbf{Algorithm 1:} Fit non-parametric ERGM, i.e. estimate $\vecbeta = \bigl( \theta_0, \vecu^t \bigr)^t$ and $\veclambda$.
\hrule
\textit{Preparation}: Fit Standard GLM for 
\begin{align}
\text{logit}\left[ \Prob\bigl( Y_{ij} = 1 | \bY \backslash Y_{ij} ; \vectheta\bigr) \right] = \theta_0 + \sum \limits_{l = 1}^{p} \Delta_{ij} s_l(\by) \theta_l \notag
\end{align}
to determine effect directions. The smooth effect $m_l(\cdot)$ is constrained to \begin{enumerate}
\item[(a)] a monotonically increasing function if $\widehat{\theta}_l \geq 0$, and
\item[(b)] a monotonically decreasing function if $\widehat{\theta}_l < 0$.
\end{enumerate} 
Matrix $\boldsymbol{A}$ is set up using the resulting monotonicity constraints according to \eqref{eq:monoton}. \\
\hrule
Instead of maximizing $\lik_p\bigl( \vecbeta, \veclambda \bigr) = \lik_p\bigl( \vecbeta, \veclambda \bigr)$ directly under the constraints from \eqref{eq:monoton}, we use a Taylor expansion of
\[
\lik_p\bigl( \vecbeta, \veclambda \bigr) - \lik_p\bigl( \vecbeta^{(t)}, \veclambda \bigr) 
\approx \vecs_p\bigl( \vecbeta^{(t)}, \veclambda \bigr)^t \bigl( \vecbeta - \vecbeta^{(t)} \bigr) + \frac{1}{2} \bigl( \vecbeta - \vecbeta^{(t)} \bigr)^t \matH_p\bigl( \vecbeta^{(t)}, \veclambda \bigr) \bigl( \vecbeta - \vecbeta^{(t)} \bigr),
\]
where $\vecs_p (\cdot)$ denotes the penalized score function and $\matH_p$ the penalized Hessian. \vspace*{12pt}\\
\textit{Initiate} starting values $\vecbeta^{(0)}$, $\veclambda^{(0)}$, $t = 0$, $s = 0$.
\begin{enumerate}
\renewcommand{\labelenumi}{\textit{Step \arabic{enumi}:}}
\item Use current value $\veclambda^{(s)}$ and iterate until convergence or until max. no. of iterations $t_{\max}$ is reached:
	\begin{enumerate}[(i)]
		\item Solve $\min_{\vecb} \left(-\vecd^t \vecb + 1/2 \vecb^t \matD \vecb \right)$ for $\vecb = \bigl( \vecbeta - \vecbeta^{(t)} \bigr)$ with constraint $\matA^t \vecb \geq \vecb_0$, \\
		where $\vecd = \vecs_p\bigl( \vecbeta^{(t)}, \veclambda^{(s)} \bigr)$, $\matD = - \matH_p\bigl( \vecbeta^{(t)}, \veclambda \bigr)$ and $\vecb_0 = - \matA^t \vecbeta^{(t)}$.
		\item Update $\vecbeta^{(t + 1)} = \vecbeta^{(t)} + \vecb$.
		\item Set $t = t + 1$. 
	\end{enumerate}
\item As long as maximum no. of iterations $s_{\max}$ or convergence is not reached:
	\begin{enumerate}[(i)]
		\item Use current value $\vecbeta^{(t)} = \bigl( \theta_0^{(t)}, \bigl( \vecu^{(t)} \bigr)^t \bigr)^t$ and update to $\veclambda^{(s + 1)}$ element-wise according to equation \eqref{eq:schall_main}:
		
\[
\widehat{\lambda}_l^{(s + 1)} = \frac{\mbox{tr} \left\lbrace \bigl[ \widetilde{\matF}^{-1} \bigl( \vecu^{(t)}, \veclambda^{(s)} \bigr) \widetilde{\matF}\bigl( \vecu^{(t)}, 0 \bigr) \bigr]_l \right\rbrace }{\bigl( \vecu_l^{(t)} \bigr)^t \vecu_l^{(t)}},\quad \text{for}\ l = 1, \ldots, p.
\] 
		\item Set $s = s + 1$.
		\item Set $\vecbeta^{(0)} = \vecbeta^{(t)}$ and $t = 0$ and start again with \textit{Step 1}.
		\end{enumerate}
\end{enumerate}
\hrule
\textit{Additional Steps:} \vspace{0.2cm}\\
When during fitting one of the penalty parameter $\lambda_l$ tends to infinity (in \textit{Step 2} (i)) we set the corresponding smooth effect $m_l(\cdot)$ to zero, i.e. $\widehat{m}_l(\cdot) \equiv 0$, and estimate the remaining components in the model with the above procedure. \vspace{0.2cm}\\
If in a later iteration than the first iteration solving \textit{Step 1} (i) fails (e.g., due to numeric problems), we take the current values $\vecbeta^{(t)}$ and check whether all of the estimated effects $\widehat{m}_l(\cdot)^{(t)}$, $l = 1, \ldots, p$ exceed a specified threshold in absolute value. \\
If not, i.e. at least one effect estimate is close to zero, we set the smallest smooth effect to zero and continue.
\end{framed}

\subsection{Combining the Sample Estimates}
\label{subsec:combinonparametric}
Let us now bear in mind that the estimation described in the previous subsection holds for one sample of the network. For each sample we obtain an estimate $\widehat{m}_{l<k>}(x_l) = \vecB(x_l)\widehat{\vecu}_{l<k>}$ with obvious definition for $\widehat{\vecu}_{l<k>}$. We now need to combine the sample estimates which are mutually dependent. One possible approach for combining the sample estimates would be to calculate a mean curve by just averaging the resulting parameter estimates over all samples. We follow a different path here originating in functional data analysis and compute a median curve. The median curve is more robust against outliers than the mean curve. We employ the methods developed by \cite{Sun-etal:2012} which are available in the \texttt{R} package \texttt{fda} \citep{fda}. There is a huge number of options available for computing functional depth, ranking curves accordingly, and determining a median curve (see, e.g., \citealp{Lopez-PintadoRomo:2009}, and \citealp{MoslerPolykova:2012}). We decided to use the approach of \cite{Sun-etal:2012} because it is quite fast even for a large number of curves, which seems important in our case (we use the option \texttt{"Both"} from the \texttt{fbplot} function \texttt{fda}, which first takes two curves for determining a band, and than computes a modified band depth in order to break ties between curves). We compute a joint median curve by sticking all estimated effects together. Computing marginal median curves per effect would be possible as well.

\section{Data Example}
\label{sec:data}

\subsection{Linear Estimation through Subsampling}
\label{subsec:ex_lin}

\begin{figure}[htb!]\centering
\includegraphics[width=0.8\textwidth]{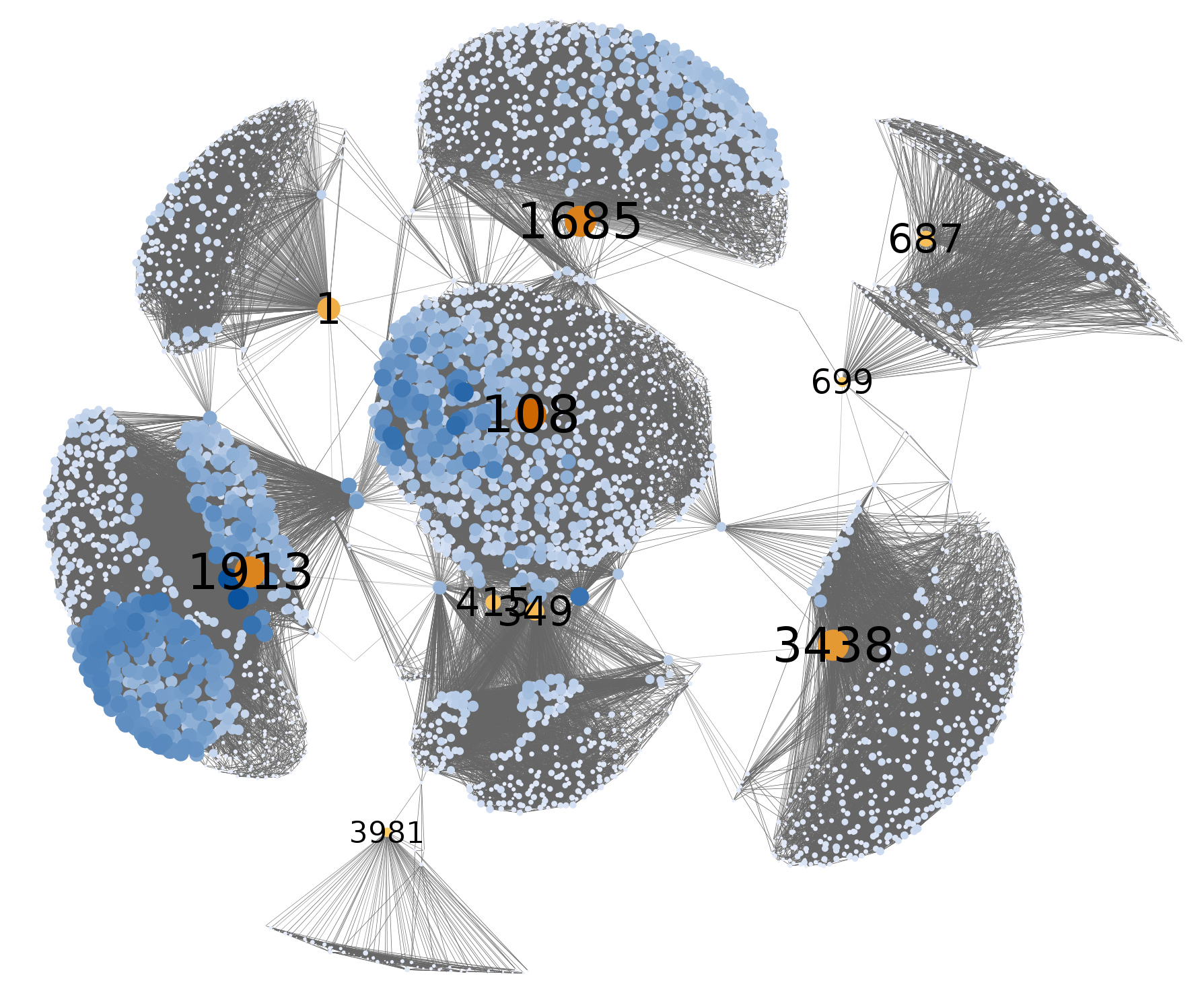}
\caption{Visualisation of the combined Facebook data. Colouring and size represent nodal degree (darker and bigger correpsonds to higher degree). Darkness and thickness of links represents the no. of triangles, the link belongs to. The ten egos are highlighted with a label indicating the node number, and coloured in orange instead of blue. Generated using stress minimization layout in visone \citep{visone}.}\label{fig:facebook}
\end{figure}

As data example we use the combined data from ten Facebook ego networks, which has originally been collected by \cite{McAuleyLeskovec:2012} and is available from the Stanford Large Network Dataset Collection \citep{snapnets}. Figure \ref{fig:facebook} shows a plot of the network graph. It is undirected and contains $88,234$ edges (Facebook friendships) between $4,039$ nodes (actors). This amounts to a network density of roughly $0.01$. 

\nocite{JuengerMutzel:2004}

\begin{figure}[htb!]\centering
\includegraphics[width=0.9\textwidth]{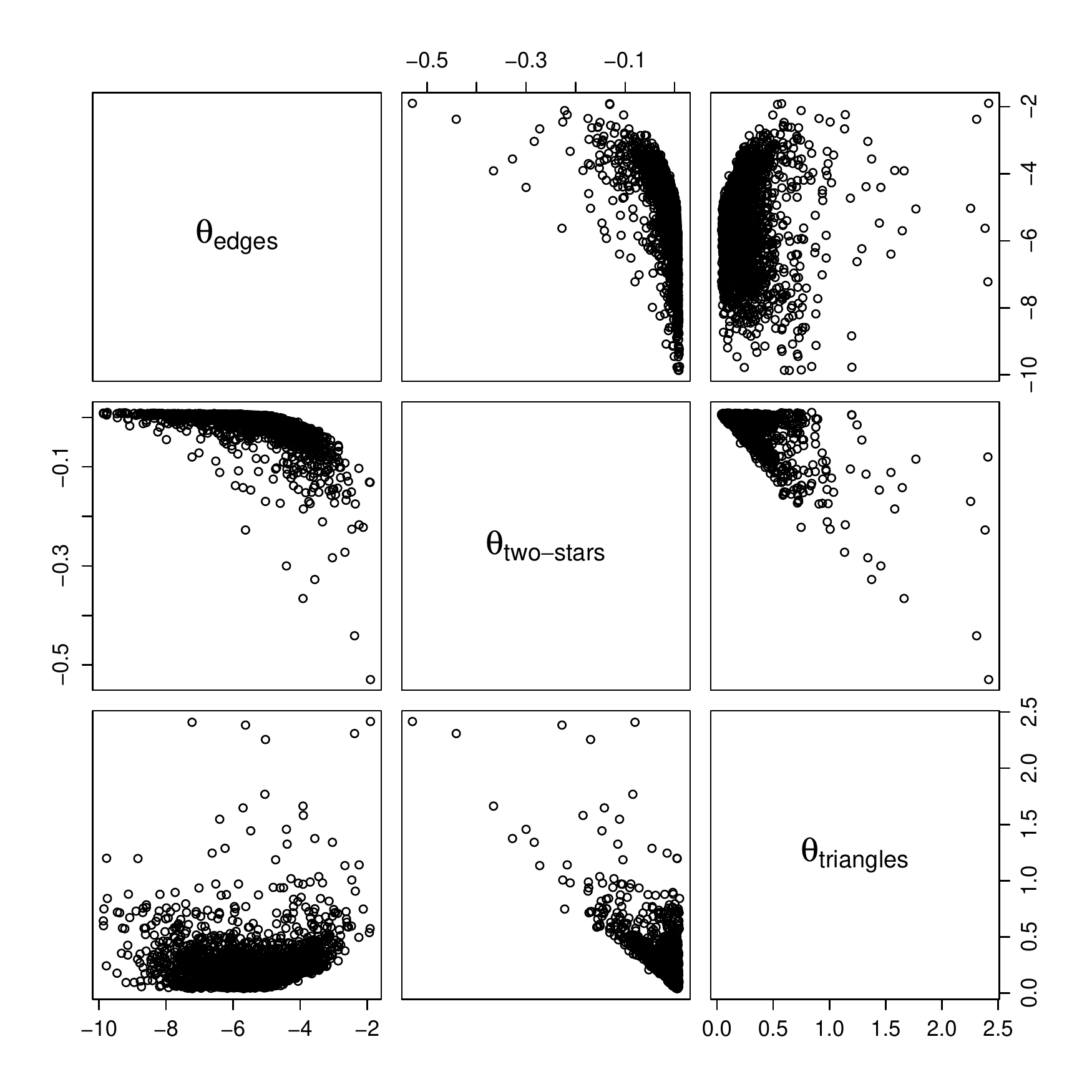}
\caption{GLM (edges, two-star, and triangle effect) results for the Facebook data. Extreme estimates with an estimated intercept $\widehat{\theta}_\text{edges} < -10$ (115 estimates) are not shown.}\label{fig:facebook_glm}
\end{figure}

We use data from the first $4,038$ rows and columns of the network adjacency matrix\footnote{As the number of nodes in the network has to be even for construction of the Latin square.} and obtain $4,037$ sample subsets $\mathcal{D}_k(n|2)$. To each of these subsets we fit a standard logistic model with edges (as intercept), two-star, and  triangle effect. We exclude subsets from the analysis which contain less than three observations with $y_{ij} = 3$ (this affects 56 data subsets). Figure \ref{fig:facebook_glm} shows pairwise scatterplots of the estimated coefficients. Extreme results with an estimated intercept $\widehat{\theta}_\text{edges} < -10$ (115 estimates) are excluded. The general impression for the shown results is that the estimated triangle effect $\widehat{\theta}_\text{triangles}$ is always positive. The estimated two-star effect $\widehat{\theta}_\text{two-stars}$ is closer to zero with some positive and some negative values. There is some negative correlation between the two parameters. Table \ref{tab:facebook_glm} displays a numerical summary of the results.

\begin{table}[htb!]\centering\small
\caption{\label{tab:facebook_glm} GLM (edges, two-star, and triangle effect) results for the Facebook data. Extreme estimates with an estimated intercept $\widehat{\theta}_\text{edges} < -10$ (115 estimates) are not considered.}
\begin{tabular}{lrrrr}
\hline\\[-1ex]
Parameter & mean est. & median est. & 5\% quantile & 95\% quantile \\ [1ex]
\hline\\[-1ex]
$\theta_\text{edges}$ & $-5.436$ & $-5.425$ & $-7.373$ & $-3.687$ \\[1ex]
$\theta_\text{two-stars}$ & $-0.012$ & $-0.003$ & $-0.054$ & $0.006$ \\[1ex]
$\theta_\text{triangles}$ & $0.207$ & $0.174$ & $0.063$ & $0.483$ \\[1ex]
\hline\\[-1ex]
\end{tabular}
\end{table}
Apparently, in a network of this size we are faced with degeneracy using a two-stars and triangles as model statistics. We therefore do not put too much emphasis in the analysis of the parametric model but go forward to a non-parametric approach in the next subsection.

\subsection{Non-parametric Estimation through Subsampling}
\label{subsec:ex_np}

We stick to the Facebook data example and continue our analysis with a non-parametric Exponential Random Graph Model as described in Section \ref{sec:nonparametric}. We fit a model containing the edge effect (as intercept), a smooth two-star effect, and a smooth triangle effect. \\
We exclude subsets from the analysis which contain less than 10 observations with $y_{ij} = 1$, because otherwise we do not have enough information for a stable estimation of the smooth effects. This affects $577$ out of $4,037$ data subsets.
We use $20$ exponential distributions as basis functions for each smooth component, with parameters $\gamma_q$ ranging between values of $0.0005$ and $1$ as displayed in Figure \ref{fig:basis}.
The maximum number if iterations is $20$, which is not reached for any of the fits. The convergence criterion is set to $1\mathrm{e}{-12}$ and for $83$ samples the algorithm is aborted, which, e.g., may be caused by separability in these subsets and resulting in non-identifiability. As described in the algorithm in the previous section, the fitted model can simplify if the fitted $\lambda_l$ goes to infinity. In this case the corresponding functional fit $\widehat{m}_l(\cdot)$ equals zero and the model is reduced. This implies that the fitting algorithm itself conducts a model selection. In addition, effects can be set to zero if numerical issues occur when solving the quadratic problem in \textit{Step 1} of the algorithm and the current effect estimate $\widehat{m}_l(\cdot)$ is close to zero (we use a value of $0.005$ for this criterion here). We therefore record the number of samples where the algorithm converges to a simplified model. Let the different models be labelled as follows:
\[
\begin{array}{lll}
M_1: & \text{``two-star''}\ \ + &\text{``triangle''} \\
M_2: &                        &\text{``triangle''} \\
M_3: & \text{``two-star''} & \\
M_4: & \multicolumn{2}{l}{\text{intercept only}}  \\
\end{array}
\]  
The notation means that model $M_2$, for instance, corresponds to a model where the non-parametric smooth two-star effect is set to zero, while for model $M_4$ both, two-star and triangle effect are set to zero. Table \ref{tab:facebook_np} summarises the results numerically and shows the number of samples for the converged models. There is a clear dominance for model $M_2$ with intercept and smooth non-parametric triangle effect only.

\begin{table}[htb!]\centering\small
\caption{\label{tab:facebook_np} Numeric summary of the model fitting results for the Facebook data. The non-parametric ERGM contains edges, a smooth two-star, and a smooth triangle effect.}
\begin{tabular}{lcr}
\hline\\[-1ex]
Total no. of samples available: & & $4,037$ \\[1ex]
\hline\hline\\[-1ex]
No. of samples with no fit (less than 10 times y = 1 in sample): & & $577$ \\[1ex]
\hline\\[-1ex]
Model $M_1$: & & $181$ \\[1ex]
Model $M_2$: & & $3,189$ \\[1ex] 
Model $M_3$: & & $5$ \\[1ex]
Model $M_4$: & & $2$ \\[1ex]
\hline\\[-1ex]
No. of samples where max. no. iterations was reached: & & $0$ \\[1ex]  
No. of samples with other reason for non-convergence: & & $83$ \\[1ex]
\hline\\[-1ex]
\end{tabular}
\end{table}

\begin{figure}[htb!]\centering
\includegraphics[width=0.9\textwidth]{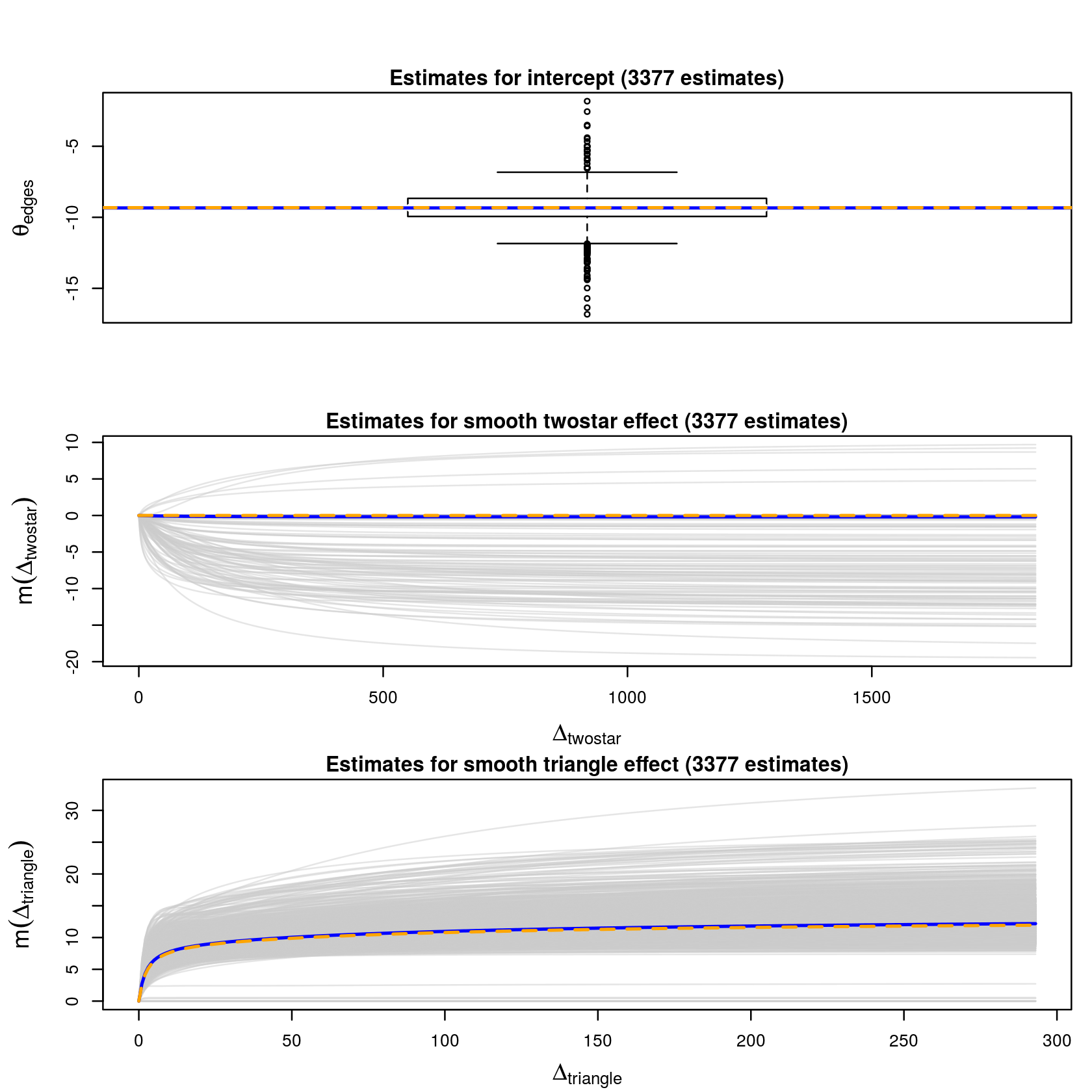}
\caption{Non-parametric ERGM (edges, smooth two-star, and smooth triangle effect) results for the Facebook data. $3,377$ estimates with convergence or effects set to zero are shown. The blue solid lines show the mean estimate; the orange dashed lines depict the median estimate.}\label{fig:facebook_smooth}
\end{figure}

Figure \ref{fig:facebook_smooth} shows the resulting $3,377$ estimates (subsets with convergence, or effect set to zero), containing mean (solid blue lines) and median estimates (dashed orange lines). As general impression we obtain a negative intercept, a positive triangle effect, and a two-star effect which is set to zero for most samples and fluctuates around zero in the remaining cases. The median curve for the two-star effect is exactly zero. 

\begin{figure}[htb!]\centering
\begin{tabular}{c}
\includegraphics[width=0.9\textwidth]{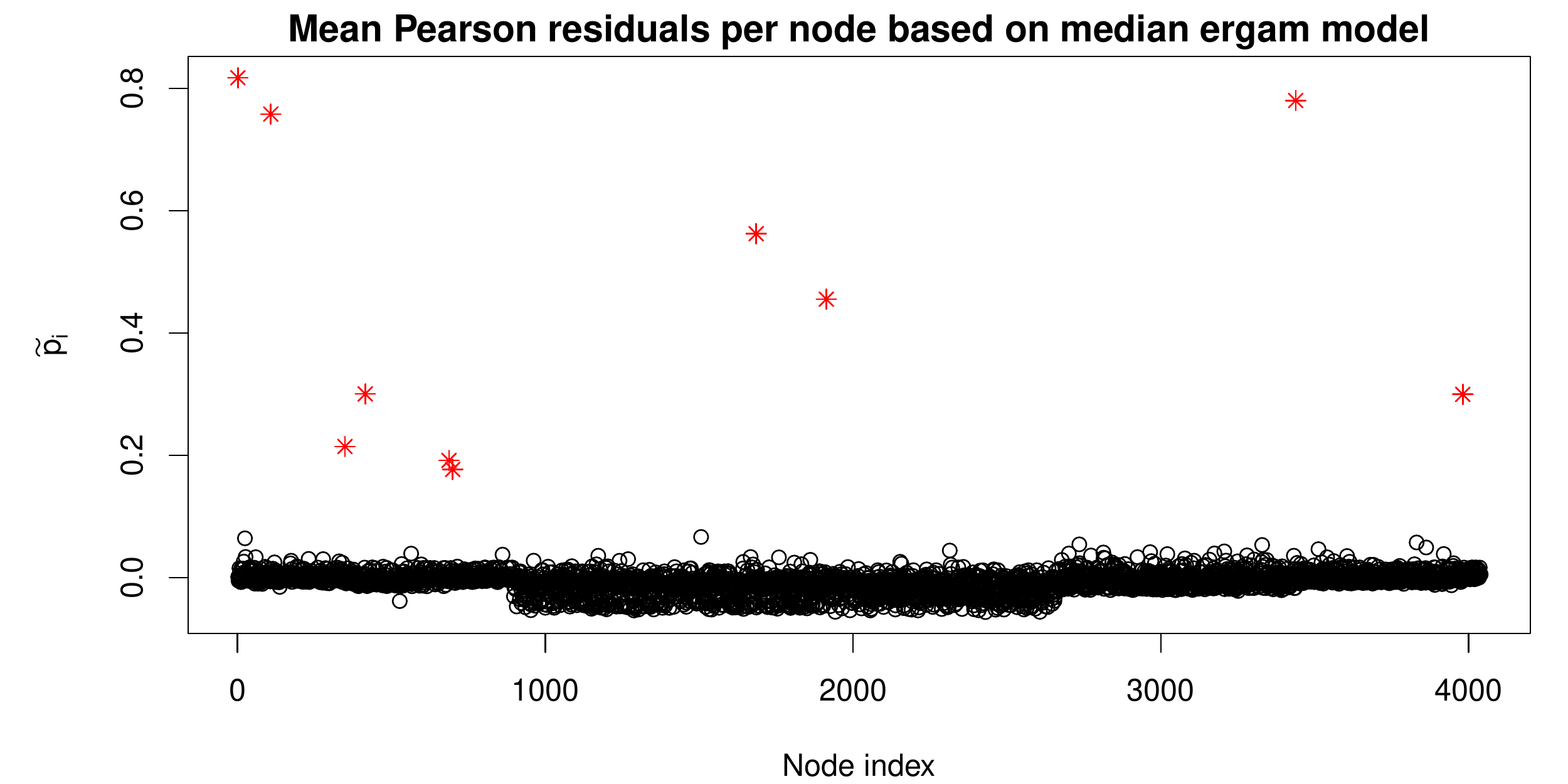} \\
\includegraphics[width=0.9\textwidth]{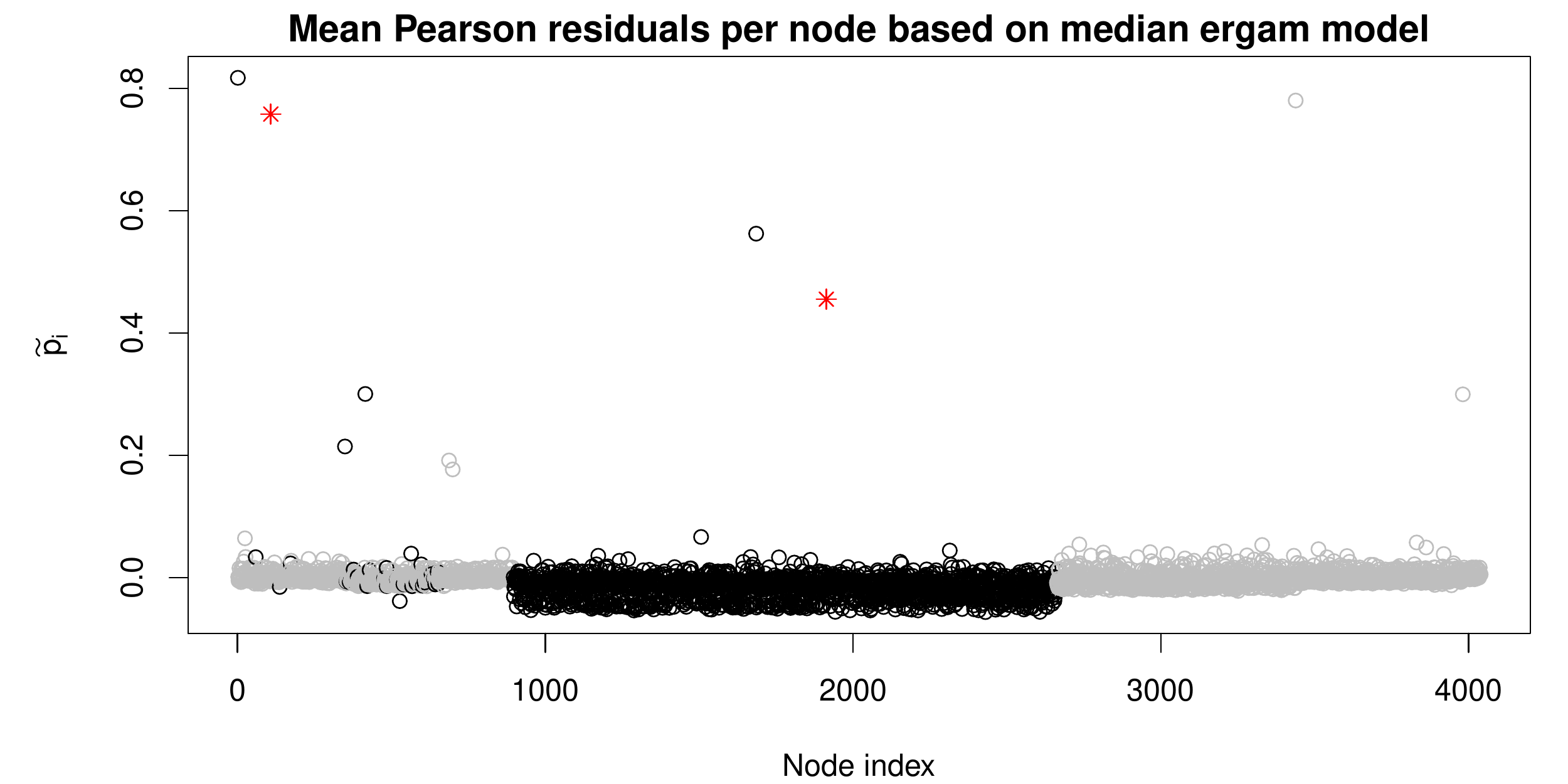} \\
\end{tabular}
\caption{Node-specific average Pearson residuals from non-parametric ERGM for the Facebook data. Prediction for the residuals is based on the overall median model. The ten egos in the network are denoted with a red star in the upper plot.
The lower plot highlights to ego-nets (for nodes nodes $108$ and $1913$). The two egos are denoted with a red star, the corresponding members of their ego-networks are black, the remaining ones of the whole dataset grey.}\label{fig:facebook_residuals}
\end{figure}

To explore the validity of the model, we continue our analysis by computing Pearson residuals for all observations
\[
e_{ij} = \frac{y_{ij} - \widehat{\pi}_{ij} }{ \sqrt{ \widehat{\pi}_{ij} ( 1 - \widehat{\pi}_{ij} ) } }, \quad \text{for}\ i = 1, \ldots, n,\ j = 1, \ldots, n,\ i \neq j, 
\]
where $\widehat{\pi}_{ij}$ is a prediction based on the obtained median (curve). As next step we calculate the average Pearson residual for each node through
\[
\widetilde{e}_i = \frac{1}{n - 1} \sum\limits_{j \neq i} e_{ij},\quad \text{for}\ i = 1, \ldots, n.
\]
Figure \ref{fig:facebook_residuals} shows the resulting node-specific average Pearson residuals. Bear in mind that the residuals are not independent, as we are averaging over non-independent samples. They should still have an expected value of zero. Figure \ref{fig:facebook_residuals} shows a clear structure. The nodes with large average residuals $\widetilde{e}_i$ are not surprising, as these are the ten egos from the network construction. In the upper plot all ten egos are depicted with a red star. They have more connections as one would expect from the model and therefore stick out. Moreover, some nodes have rather negative Pearson residuals and these nodes can be attributed to specific ego-nets. The lower plot in Figure \ref{fig:facebook_residuals} highlights two ego-nets (for nodes $108$ and $1913$, these two egos are again depicted with a red star, the residuals belonging to the ego-nets are black, the remaining ones grey) and they account for almost all of these negative residuals. Our conclusion from this residual analysis is that for some parts of the network the overall model seems too simplistic, while for others the outcome appears to be reasonable.

\clearpage

\begin{figure}[htb!]\centering
\includegraphics[width=0.9\textwidth]{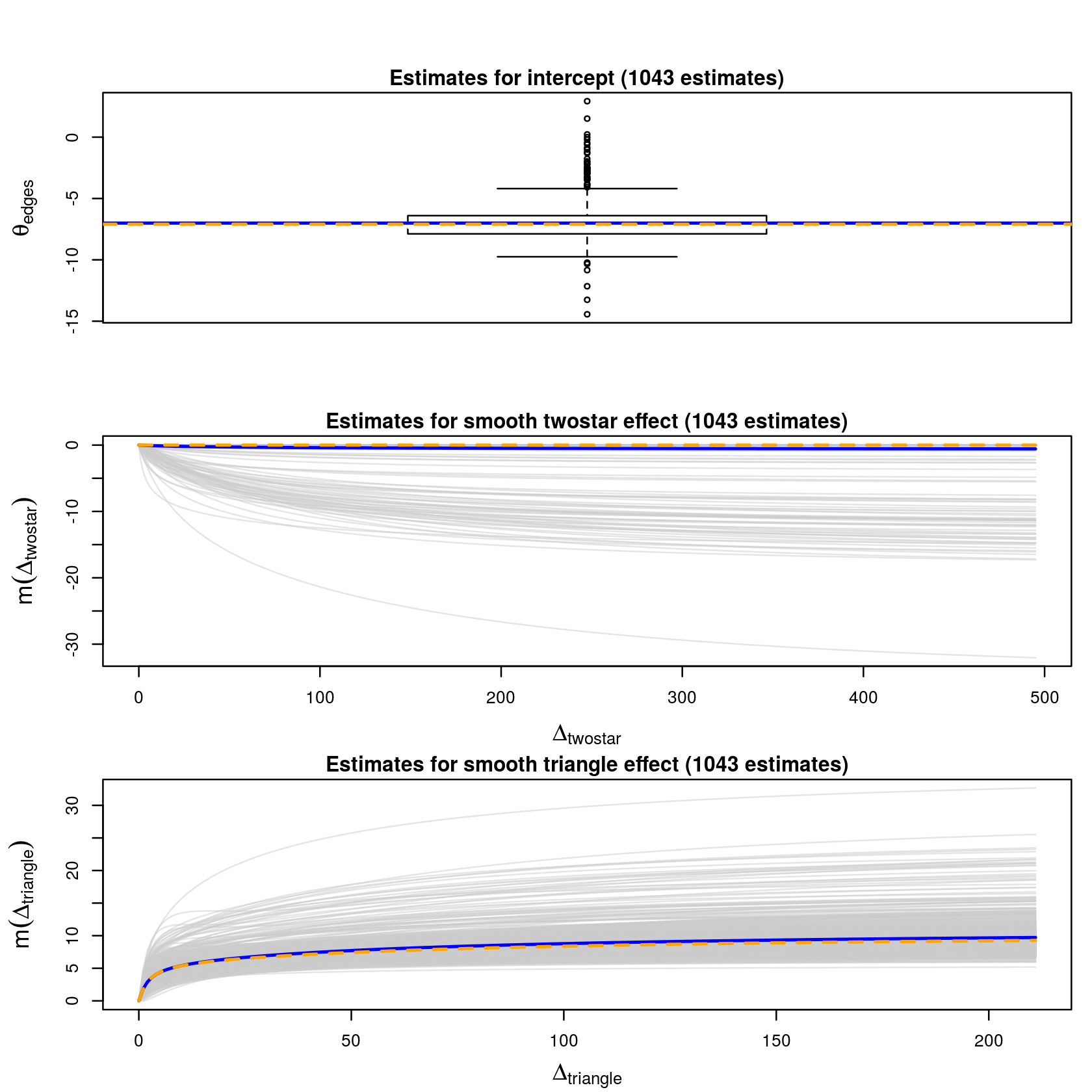}
\caption{Non-parametric ERGM (edges, smooth two-star, and smooth triangle effect) results for ego-net of node $108$ from the Facebook data. $1,043$ estimates with convergence or effects set to zero are shown. The blue solid lines show the mean estimate; the orange dashed lines depict the median estimate.}\label{fig:facebook_smooth_108}
\end{figure}

\begin{figure}[htb!]\centering
\includegraphics[width=0.9\textwidth]{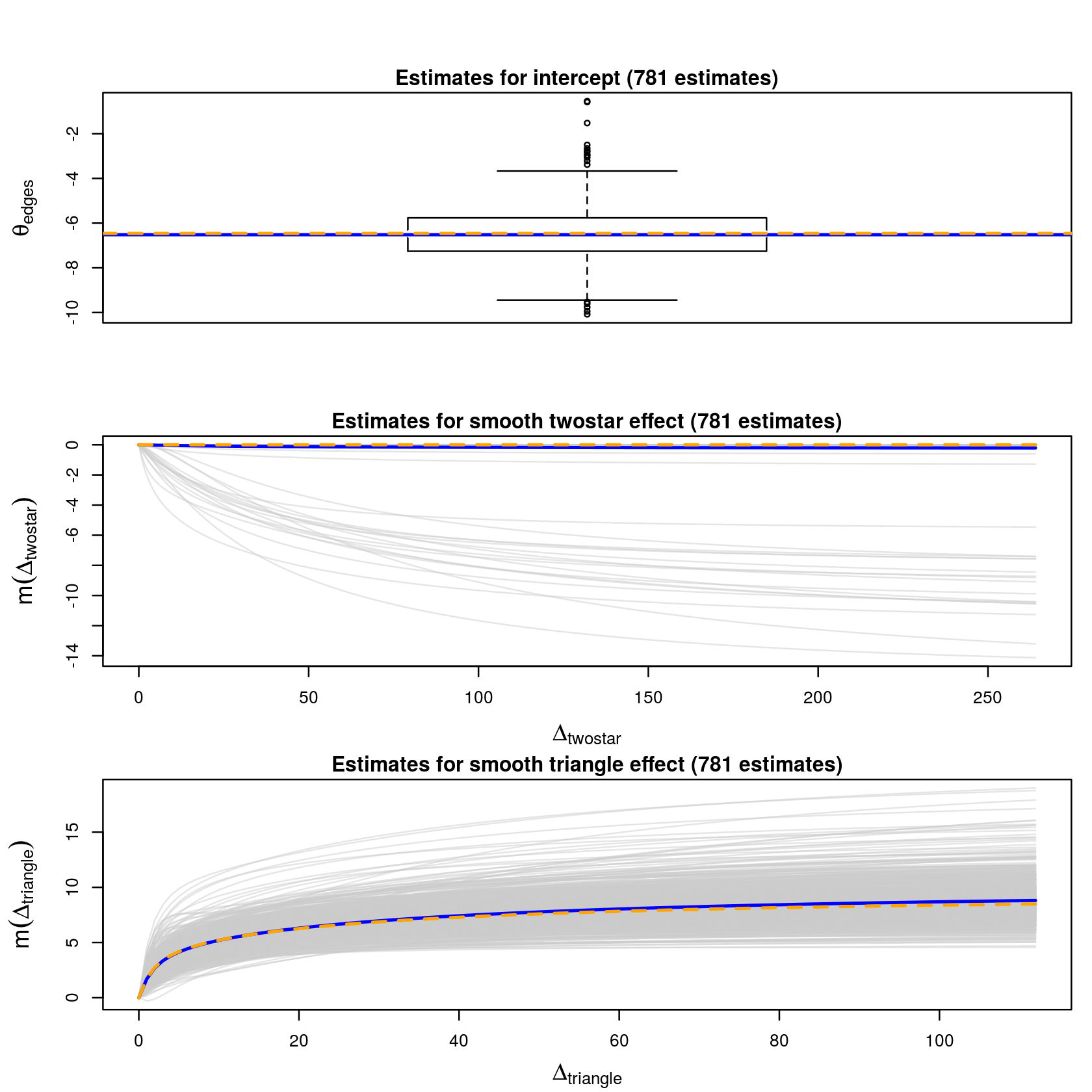}
\caption{Non-parametric ERGM (edges, smooth two-star, and smooth triangle effect) results for ego-net of node $1,685$ from the Facebook data. $781$ estimates with convergence, or effects set to zero are shown. The blue solid lines show the mean estimate; the orange dashed lines depict the median estimate.}\label{fig:facebook_smooth_1685}
\end{figure}

We continue our analysis and look at the ego-nets of node $108$ (contains a majority of nodes with negative average Pearson residual; consists of $1,045$ nodes with $26,750$ edges), and of node $1,685$ (consists of $792$ nodes with $14,025$ edges) separately. The egos themselves are not part of the subnetworks as they are connected to every other vertex in the corresponding subnetwork (by construction). The setup for the fit is the same as before for the non-parametric ERGM. Figures \ref{fig:facebook_smooth_108}, and \ref{fig:facebook_smooth_1685} show the corresponding estimates. Table \ref{tab:facebook_np_sub} summarises the results. When comparing the results to the ones for the whole dataset, the overall impression is similar, with a positive triangle effect, and a two-star effect close to zero (or set to zero for most samples, and a zero median curve). The intercept values are smaller in absolute value, which is not surprising as we are analysing smaller networks (with a higher density).

\begin{table}[htb!]\centering\small
\caption{\label{tab:facebook_np_sub} Numeric summary of the model fitting results for ego-subnets of the Facebook data. The non-parametric ERGM contains edges, a smooth two-star, and a smooth triangle effect.}
\begin{tabular}{lcrcr}
Ego-net && $108$ && $1,685$ \\ 
\hline\hline\\[-1ex]
Total no. of samples available: & & $1,043$ & & $791$ \\[1ex]
\hline\hline\\[-1ex]
No. of samples with no fit (less than 10 times y = 1 in sample): & & $132$ & & $65$ \\[1ex]
\hline\\[-1ex]
Model $M_1$: & & $0$ & & $10$  \\[1ex]
Model $M_2$: & & $911$ & & $716$ \\[1ex] 
Model $M_3$: & & $0$ & & $0$ \\[1ex]
Model $M_4$: & & $0$ & & $0$ \\[1ex]
\hline\\[-1ex]
No. of samples where max. no. iterations was reached: & & $0$ & & $0$ \\[1ex]  
No. of samples with other reason for non-convergence: & & $0$ & & $0$ \\[1ex]
\hline\\[-1ex]
\end{tabular}
\end{table}

\begin{figure}[htb!]\centering
\begin{tabular}{c}
\includegraphics[width=0.9\textwidth]{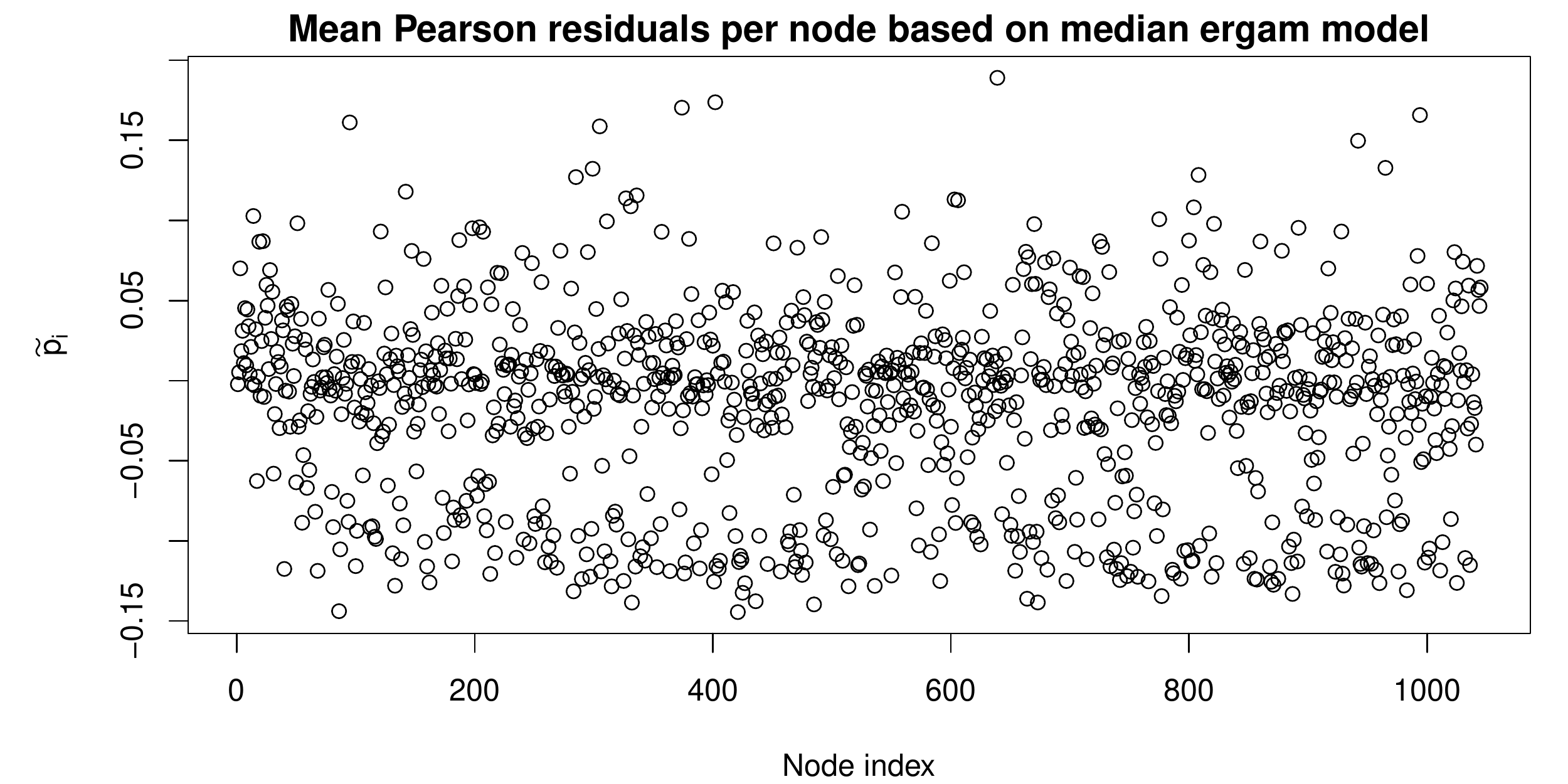} \\
\includegraphics[width=0.9\textwidth]{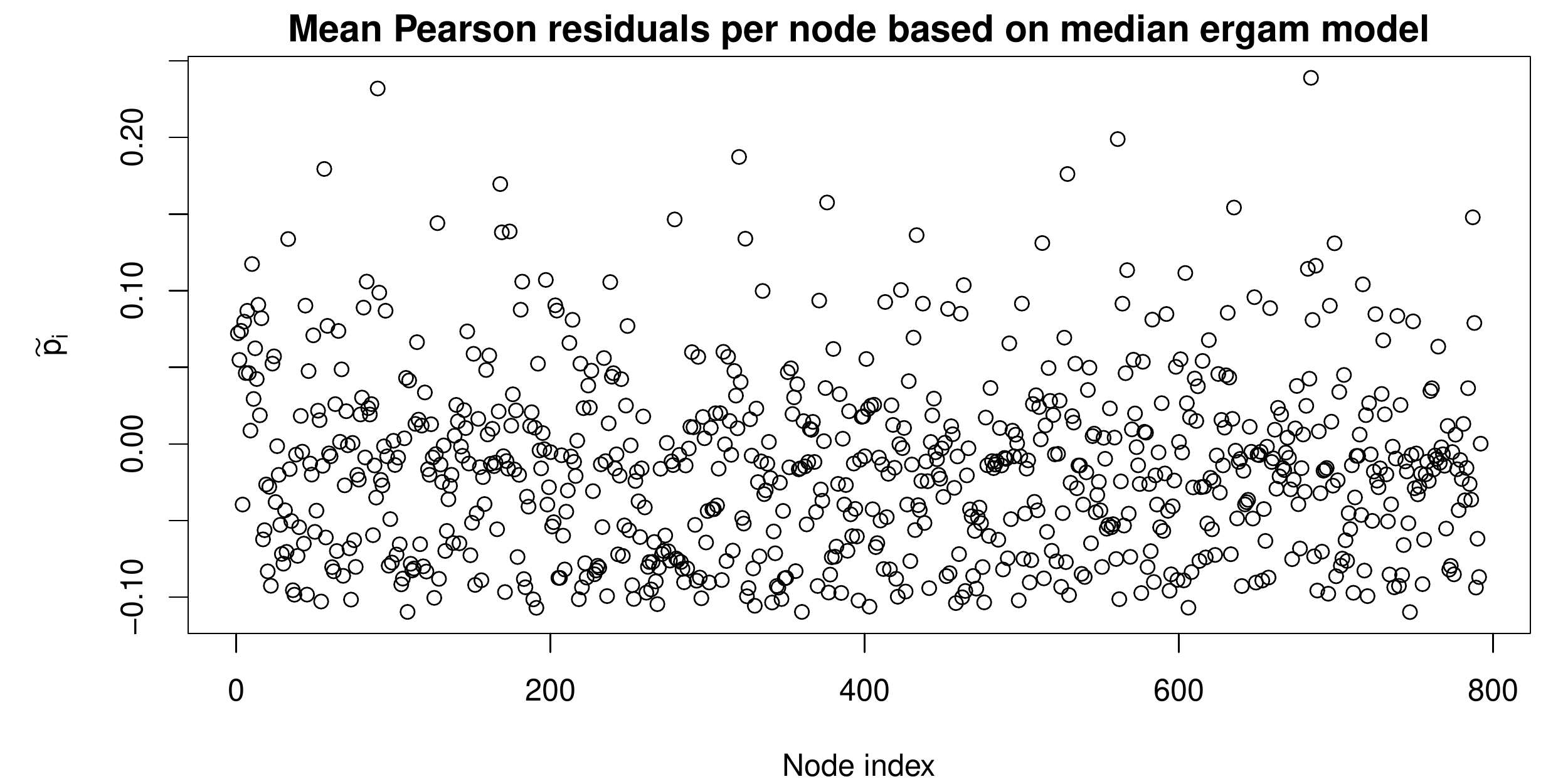} \\
\end{tabular}
\caption{Node-specific average Pearson residuals from non-parametric ERGM for the ego-nets of node $108$ (upper plot) and $1,685$ (lower plot) from the Facebook data.}\label{fig:facebook_residuals_108-1685}
\end{figure}

Figure \ref{fig:facebook_residuals_108-1685} shows the resulting nodes-specific average Pearson residuals for both ego-nets. The result looks more homogeneous than before, but for the ego-net of $108$ we still see that there are nodes in the network with a rather negative average Pearson residual, i.e. they have fewer connections than the model would predict. This might be solved by extending the modelling approach and include node-specific or dyadic covariates into the model. This is of course easily possible in combination with the smooth effects but lies beyond the scope of this paper.

\clearpage

\section{Discussion}
\label{sec:discussion}

We have shown that it is possible to make use of the Markov independence assumption in the context of Exponential Random Graph Models to obtain samples consisting of independent observations which allow to use standard generalized linear models (GLM) for model fitting. Extending this approach to generalized additive models (GAM) by adding smooth functional components in a non-parametric fashion enables us to gain flexibility while maintaining the simple interpretability of statistics like two-stars and triangles. It circumvents the construction or use of more complex statistics like, e.g., geometrically weighted degree or edgewise shared partners, which also stabilise the model fitting but are very difficult to interpret. In addition, the whole estimation procedure is quite fast (much faster than the standard MCMC based routines available for ERGMs) as we are using well established model fitting routines for GLMs and GAMs, and it can easily be run in parallel as the individual sample fits can be computed independently of each other. The computation for the Facebook data example was run in parallel on 20 cores (with 2.60 GHz) and took less than four minutes (including all data pre-processing and storing the results on disk). \\
To employ the described models the network needs to be big enough (otherwise the resulting samples are too small), and what can be more problematic, the network has to be dense enough as otherwise we obtain samples consisting only of observations with $y_{ij} = 0$. The later is a general problem in real-world networks, as it is well-known that with increasing network size $n$ the density tends to become smaller and smaller. The proposed modelling strategy therefore clearly has some caveats. Also, it is difficult to give a general advice on how many actors are needed for our method to work. When using the GLM approach on each subsample of size $\frac{n}{2}$ a smaller number of observations is reasonable than for the non-parametric GAM approach. In the data example we have presented in the previous section, the sample size itself is not an issue with 2,019 observations per subsample for the whole network, and 396 or 522, respectively, for the ego-nets, where we fit models with two smooth functional components plus an intercept term. Still, we had some problems with obtaining samples with enough $y_{ij} = 1$ observations per sample. If course this issue becomes more severe when the network density (which is 0.01 for the complete Facebook data, and therefore quite high for a network of this size) goes down. \\
Another problem which is apparent from the residual analysis in Figures \ref{fig:facebook_residuals}, and \ref{fig:facebook_residuals_108-1685} is that the residuals are quite low in absolute value. This is a sign or underdispersion in the underlying binomial models and can be explained by zero-inflation, i.e. we have more zeros in the data than we would expect under the model. This result is not surprising, again due to the low density in large networks, where Exponential Random Graph Models tend to be problematic in general. There are approaches going into the direction of assuming local dependence structures, whereas the standard ERGM assumes a global rather strict dependence structure and is therefore probably unrealistic especially in the context of large networks. \cite{SchweinbergerHandcock:2015} use hierarchical ERGMs, see also the corresponding \texttt{R} package \texttt{hergm} \citep{hergm}, where the neighbourhood structure can be taken into account if it is known, or estimated as a latent construct using a Bayesian approach. The later is computationally very problematic and rather time consuming or even infeasible for large networks. Another possible solution to handle the zero-inflation using our subsampling approach would be the use of mixture models as available, e.g., in the \texttt{R} package \texttt{flexmix} \citep{Leisch:2004}, and employ zero-inflation models \citep[Section 5.1]{GruenLeisch:2008} for binomial data to each sample. To us this appears to be a promising field for future research.

\subsubsection*{Acknowledgements}
We gratefully acknowledge the help of Ulrik Brandes with the visualisation of the Facebook data example in Figure \ref{fig:facebook} using visone (version 2.16).

\clearpage

\bibliography{LaTeX/literatur}

\end{document}